\documentclass[prl,twocolumn,showpacs]{revtex4}
\usepackage{graphics}\usepackage{graphicx}\usepackage{epsfig} \usepackage{amssymb}\usepackage{amsmath}

\begin{document}

\title{Laser-cluster interaction: x-ray production by short laser pulses}%

\author{Cornelia Deiss}
\email{cornelia@concord.itp.tuwien.ac.at}
\affiliation{Institute for Theoretical Physics, Vienna University of Technology, A-1040 Vienna, Austria, EU}
\author{Nina Rohringer}
\affiliation{Institute for Theoretical Physics, Vienna University of Technology, A-1040 Vienna, Austria, EU}
\author{Emily Lamour}
\affiliation{INSP, Universit\'es Paris 6 et 7, Campus Boucicaut, 75015 Paris, France, EU}
\author{Christophe Prigent}
\affiliation{INSP, Universit\'es Paris 6 et 7, Campus Boucicaut, 75015 Paris, France, EU}
\author{Jean-Pierre Rozet}
\affiliation{INSP, Universit\'es Paris 6 et 7, Campus Boucicaut, 75015 Paris, France, EU}
\author{Dominique Vernhet}
\affiliation{INSP, Universit\'es Paris 6 et 7, Campus Boucicaut, 75015 Paris, France, EU}
\author{Joachim Burgd\"orfer}
\affiliation{Institute for Theoretical Physics, Vienna University of Technology, A-1040 Vienna, Austria, EU}

\date{\today}
\begin{abstract}
We investigate the heating of the quasi-free electrons in large rare gas clusters ($N$ exceeding $10^5$ atoms) by short laser pulses at moderate intensities ($I\simeq 10^{15}\,\mathrm{Wcm^{-2}}$). We identify elastic large-angle backscattering of electrons at ionic cores in the presence of a laser field as an efficient heating mechanism resembling the Fermi shuttle. Its efficiency as well as the effect of collective electron motion, electron-impact ionization and cluster charging, are studied employing a mean-field classical transport simulation. Results for the absolute x-ray yields are in surprisingly good quantitative agreement with recent experimental results.
\end{abstract}
\maketitle

The interaction of short and ultra-short intense laser pulses with clusters has become an important area of laser-matter research bridging the gap between gas-phase and solid-state processes \cite{krain}. The observation of characteristic x-ray emission from laser-irradiated clusters \cite{paris1,parishci,paristhese,parisprl} suggested its potential as an x-ray source through a highly non-linear conversion of IR radiation that combines advantages of both solid and gaseous targets. Like solids, clusters provide large x-ray yields, yet unlike solids they are debris-free, just like gaseous targets. 
Characteristic x-ray emission also provides important time-differential information on the laser-induced electronic dynamics on a femtosecond scale. The charge state as well as the vacancy distribution of the cluster ions at the instant of emission can be extracted from high-resolution x-ray spectra. Moreover, as the vacancy production in deeply bound shells (e.g.\ K-shell in argon or L-shell in xenon) proceeds via impact ionization by energetic electrons, characteristic x-rays provide an ``in situ'' thermometer of the temperature of the heated electron gas in the cluster. Recent experiments \cite{parishci,paristhese,parisprl} found an unexpectedly low laser intensity threshold for the production of x-ray radiation. When irradiating large argon clusters with $N > 10^5$ atoms with laser pulses with a short pulse duration of $\tau=60\,\mathrm{fs}$ at full width half maximum (FWHM), the intensity threshold for the production of characteristic K-x-rays lies at $I_{th}=2.2\cdot10^{15}\,\mathrm{Wcm^{-2}}$. By comparison, the ponderomotive energy, $U_P=F^2/(4\omega^2)$, of a free electron in a laser field of this intensity is as low as $U_P\simeq130\,\mathrm{eV}$, more than an order of magnitude below the binding energy $E_K\simeq3.2\,\mathrm{keV}$ of the K-shell electrons. This observation raises puzzling questions as to the efficient heating mechanism for electrons in large clusters at such moderate intensities of very short pulses with $\sim 40$ optical cycles. 

\begin{figure}
\includegraphics[angle=90,width=7cm]{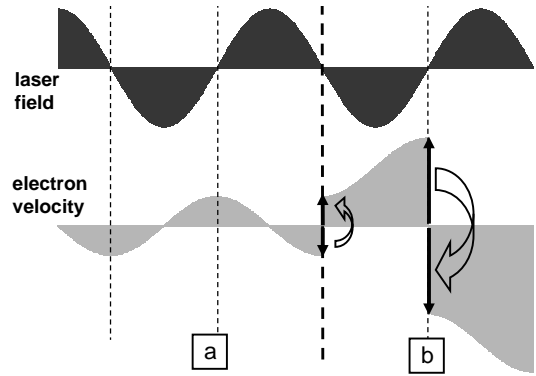}
\caption{\label{fig:1} Time evolution of the velocity of an electron in a laser field, schematically. A free electron does not experience effective velocity gain beyond the quiver velocity (a). If the velocity is flipped in synchrony (b), an electron can rapidly absorb kinetic energy from the laser field.}
\end{figure}
\begin{figure}[tb]
\includegraphics[angle=90,width=8.5cm]{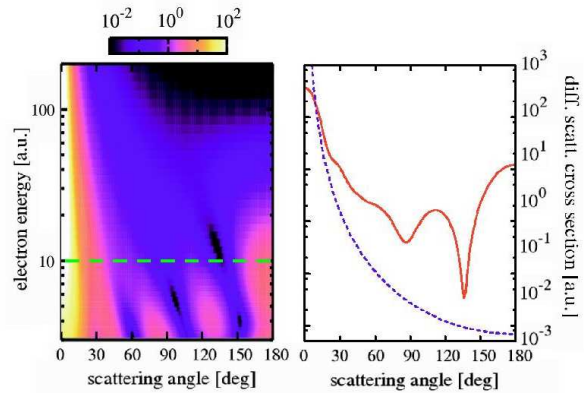}
\caption{\label{fig:2} (Color online) Left: Differential cross-section distribution $d\sigma_e(\theta,E)/d\Omega$ in $\mathrm{a.u.}$ for elastic electron scattering at $\mathrm{Ar}^{2+}$ ions. The cross section was calculated by partial wave analysis for a parameterized Hartree-Fock potential \cite{szydlik,salvat}.
Right: $d\sigma_e(\theta)/d\Omega$ for an electron with fixed kinetic energy $E=10\,\mathrm{a.u.}$ (solid line). For comparison, the Rutherford cross-section $\left(\frac{q}{4E}\right)^2\frac{1}{\sin^4(\theta/2)}$ is also displayed (dashed line).}
\end{figure}

Several theoretical models for intense laser-cluster interaction have been proposed \cite{krain,ditmire,last1,rose,saalmann,brabec,smirnov}, none of which appears to provide a satisfactory explanation for such rapid acceleration of electrons. A theoretical description of intense laser-cluster interaction represents a considerable challenge in view of the many-body nature of this process. Molecular dynamic simulations \cite{last1,rose,saalmann} are limited to about 1000 atoms, and results obtained for small clusters are difficult to scale to larger sizes. The recently proposed microscopic particle in cell (MPIC) method \cite{brabec} reaches clusters of $\sim 10^4$ atoms. Larger clusters with $N> 10^5$ particles appear still not in reach, and quantitative predictions for x-ray emission and inner-shell processes have not yet been attempted. We propose in this letter an efficient heating mechanism of electrons in large clusters that is operational within a few optical cycles and at moderate laser intensities. It is based on the observation that \emph{elastic} large-angle scattering of electrons at cluster atoms (ions) in the simultaneous presence of a laser field provides an efficient route to electron acceleration. Elastic backscattering at the core potential of the ions can flip the velocity vector of an electron, allowing it with non-negligible probability to remain synchronized with the alternating laser field vector (Fig.\ \ref{fig:1}).
 Consequently, the electron will rapidly accumulate rather than lose momentum during subsequent half cycles. Within a few optical cycles, electrons can thus be accelerated to high kinetic energies well beyond the quiver energy $E_p = 2 U_p$. This heating mechanism resembles  the Fermi shuttle acceleration \cite{fermi,burg91} and is also related to the lucky-electron model proposed for IR photoemission from metallic surfaces \cite{lucky}. The acceleration of a particle by successive backscatterings from a moving and a stationary target in the original proposal by Fermi for acceleration of cosmic particles is here modified such that the cluster ions represent stationary targets, while the alternating force field of the laser plays the role of the moving target. 
 A realistic estimate for the efficiency of this heating mechanism hinges on a proper description of the differential elastic scattering cross sections for electrons, $\mathrm{d}\sigma_e/\mathrm{d}\theta$, into backward angles $\theta \gtrsim 90^{\circ}$, which are determined by the non-Coulombic short-ranged potentials of the ionic cores.  $\mathrm{d}\sigma_e/\mathrm{d}\theta$ was calculated for electron scattering at argon ions for all relevant charge states between $q=1$ and $q=16$ over a wide range of energies using parametrized Hartree-Fock potentials \cite{szydlik,salvat}. $\mathrm{d}\sigma_e/\mathrm{d}\theta$ is typically dominated by few low-order partial waves giving rise to generalized Ramsauer-Townsend minima \cite{mott} and diffraction oscillations \cite{burg95} (Fig.\ \ref{fig:2}).
  We assumed for simplicity the electronic ground state occupation for each $q$. Extensions to core-excited configurations would be straight forward. For the interstitial region a muffin-tin potential is employed in order to account for solid-state effects \cite{salvat}. The potential shape in this region has, however, no significant influence on the cross sections at backward angles. The latter exceeds the pure Coulomb case by several orders of magnitude for all charge states and over a wide range of electron energies ($\lesssim keV$). The frequent usage in simulations of unrealistic (softened) Coulomb potentials \cite{last1,saalmann,brabec}, which grossly underestimates backscattering, is quite likely one reason why this route of electron acceleration has not yet been accounted for. Moreover, this process becomes much more important for large clusters, as the mean-free path for elastic scattering becomes comparable to the cluster size. The important role of realistic core potentials has recently also been identified in the quantum analogue of this process, inverse Bremsstrahlung ($U_P \ll \hbar \omega$), for clusters in a vacuum ultraviolet (VUV) laser field \cite{santra}. 

A full ab initio simulation for large clusters $(N \gtrsim 10^5$ particles) appears still impractical. In the following we present a simplified theoretical description of the electronic ensemble that allows to tackle its short-time dynamics ($\tau \sim 60 \, \mathrm{fs}$). It employs a generalization of classical transport theory (CTT \cite{burg90}) for open systems, in which the electronic dynamics is represented by a classical phase-space distribution $f (\boldsymbol{r}, \boldsymbol{\dot{r}}, t)$ whose evolution is determined by test-particle discretization, i.e. by solving the corresponding Langevin equation for representative trajectories. In the present case, the ensemble consists of $N_e$ quasi-free electrons, liberated inside the cluster after ionization of the cluster atoms, represented by $N_{test}=\alpha N_e$ particles, where the scaling parameter $\alpha$ is limited by computational feasibility. Each test particle is subject to a Langevin equation (atomic units are used unless otherwise stated),
\begin{eqnarray}
\label{eq:1}
\boldsymbol{\ddot{r}}_i&=&- \boldsymbol{F}_L (t) - \boldsymbol{\nabla} V (\boldsymbol{r}_i, t) \nonumber\\
&&-\boldsymbol{F}_{mean} (\boldsymbol{r}_i, t) + \boldsymbol{F}_{stoc} (\boldsymbol{r}_i, \boldsymbol{\dot{r}}_i, t)
\end{eqnarray}
with $i=1,\ldots N_{test} (t)$. Eq.\ \ref{eq:1} describes a dynamical system open to both particle number variation, $N_{test} (t)$, due to successive ionization events, and energy exchange with the many-particle reservoir (atoms, ions, and electrons) as well as with the laser field taken to be linearly polarized with a temporal envelope 
\begin{equation}
\label{eq:2}
\boldsymbol{F}_L(t)=F_0\boldsymbol{\hat{z}}\mathrm{sin}(\omega t)\mathrm{sin}^2\left(\frac{\pi t}{2\tau}\right).
\end{equation}
Eq.\ (\ref{eq:1}) provides a computational starting point for treating many-body collisional correlation effects through stochastic forces $\boldsymbol{F}_{stoc}$, which can be determined either from independent ab-initio quantum calculations or experimental data \cite{burg90}. Forces resulting from static conservative potentials $V$ inside the cluster \cite{last1}, can be included as well. Because of the coherent motion and high ionization density, the present extension of the CTT (Eq.\ \ref{eq:1}) goes beyond the independent-particle description by including dynamic electron-electron and electron-ion interactions on a time-dependent mean field level. Accordingly, $\boldsymbol{F}_{mean}$ depends on the entire ensemble of test-particle coordinates $\left\{ \boldsymbol{r}_i (t) \right\}$. 
We therefore propagate the elements of the ensemble by self-consistently coupling a given trajectory to the mean field of other trajectories running in parallel by continuously updating the forces. 

Effects of fluctuations on the electronic dynamics can be taken into account through stochastic forces which are determined from Poissonian random processes. For example, electron-ion scattering, electron-impact ionization, and core-hole excitation are determined by the probability per unit pathlength for scattering
\begin{equation}
\label{eq:3}
\lambda^{-1}_{scatt} = \frac{\mathrm{d} P_{scatt}}{\mathrm{d} x} = \sigma_{scatt} (q, E) \rho (t) \, ,
\end{equation}
controlled by the energy ($E$) and charge state ($q$) dependent integral cross-section for this process, $\sigma_{scatt}(q, E)$, and the instantaneous ionic target density $\rho (t)$ of a given charge state. Each stochastic scattering process results in ``jumps'' (classical trajectory jumps and jumps in occupation) at discrete times. A jump in momentum, $\Delta \boldsymbol{\dot{r}}$, signifies elastic scattering determined by the differential cross section, $\mathrm{d} \sigma_e (q, E, \theta)/\mathrm{d} \theta$, a simultaneous jump in test-particle number, $\Delta N_{test}$, represents ionizing collisions, and a simultaneous jump in the number of inner-shell vacancies, $\Delta N_K$, results from core-exciting collisions. The key point is that the necessary input data, $\sigma_{scatt} (q,E)$, can be determined and tabulated independently from the simulation at any desired level of sophistication. 
In the present simulation, the electron-impact ionization cross sections are determined from a modified Lotz formula \cite{paristhese,lotz} $\sigma_{i}(q,E)=A^*_q\ln(E/W^*_q)/(EW^*_q)$ (for $E\geq W^*_q$),
where the empirical parameters $A^*_q$ and $W^*_q$ were obtained by a fit to experimental ion-atom collision data \cite{paristhese,zhang}. The cluster-specific effects of suppression of the work function can be incorporated by modifying the effective work function $W^*_q$ in the presence of nearby ionic cores.

For the mean field we perform a multipole expansion keeping only the monopole and dipole terms. The monopole term is given by $\boldsymbol{F}^{(0)}_{mean} (\vec{r}, t)=\langle Q(r, t)\rangle\boldsymbol{r}/r^3 $ where $\langle Q(r,t) \rangle$ is the instantaneous charge of the sphere of radius $r$ resulting from the displacement of the ensemble of test particles relative to the ionic background. Analogously, the dipole field inside the cluster ($r<R(t)$) is $\boldsymbol{F}^{(1)}_{mean}(t)=-\boldsymbol{p}(t)/R(t)^3$,
while outside it is that of a central dipole. The dipole moment, $\boldsymbol{p}$ is determined by $\boldsymbol{p}(t)\simeq-\frac{1}{\alpha}\sum_i\boldsymbol{r}_i$, 
where the sum extends over the subset of test particles with $r_i<R(t)$.

As the ionic and electronic dynamics proceed on different time scales, the onset of cluster expansion can be taken into account through the parametric variation of the radius $R(t)$ of the uniform spherical charge background representing the ions of mass $M$ in their time-dependent mean charge state $ \langle q (t) \rangle$:
\begin{equation}
M\frac{\mathrm{d^2}R(t)}{\mathrm{d}t^2}=\frac{\langle q (t)\rangle \langle Q (R,t)\rangle}{R^2(t)} \, .
\end{equation}

\begin{figure}[tb]
\includegraphics[width=8cm]{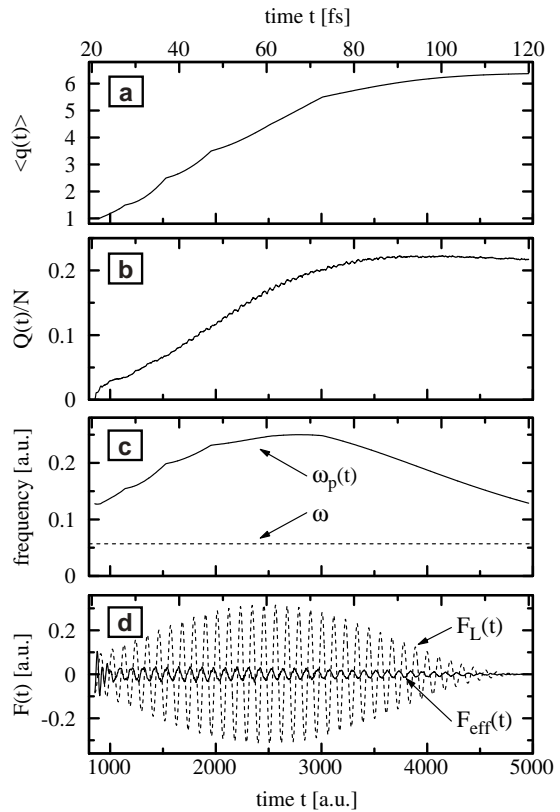}
\caption{\label{fig:3} Dynamics of an argon cluster with $N=2.8\times10^5$ atoms, irradiated by a laser pulse with wavelength $\lambda=800\,\mathrm{nm}$, duration $\tau=60\,\mathrm{fs}$, and peak intensity $I=3.5\times 10^{15}\,\mathrm{Wcm^{-2}}$. All quantities are shown as a function of time. (a) Mean ionic charge state $\langle q(t) \rangle$, (b) cluster charge per atom $Q(t)/N$, (c) comparison between plasma frequency $\omega_p(t)$ (solid line) and laser frequency $\omega$ (dashed line), (d) effective field $F_{eff}$ inside the cluster including dielectric polarization (solid line) and laser field $F_L$ (dashed line).} 
\end{figure}

We solve (Eq.\ \ref{eq:1}) for a cluster with $N = 2.8 \times 10^5$ argon atoms with initial atomic number density  $\rho(t=0)=2.66\times10^{22}\,\mathrm{cm^{-3}}$ and  initial radius  $R(0)=258 \,\mathrm{a.u.}$ irradiated by a laser pulse of length (FWHM) $\tau=60\mathrm\,{fs}$, wavelength $\lambda=\,800\mathrm{nm}$ and peak intensities  $I \gtrsim 10^{15}\,\mathrm{Wcm^{-2}}$. The ponderomotive energy of the electrons is of the order of $U_P \gtrsim 70\, \mathrm{eV} \approx 2.5 \,\mathrm{a.u.}$
As the laser field reaches for the first time the threshold field for over-barrier ionization $F_L^{1}(t_1) \simeq 0.08$, the first $N_{test} (t_{1})$ test particles with zero velocity randomly distributed over the cluster provide initial conditions for the propagation of Eq.\ (\ref{eq:1}). In the present case $N_{test} = 0.1N_e$ (or $\alpha=0.1$). Contributions from tunneling ionization can be included but are in the present case negligible. The test-particle number subsequently increases by further ionization events (Fig.\ \ref{fig:3}(a)). Impact ionization by electrons meanwhile accelerated is highly effective, making further field ionization events unlikely. The mean charge state rapidly increases to $\sim 7$, which  most likely still underestimates the ionization efficiency, as non-radiative core-hole relaxation and enhanced ionization by suppression of the work function by ion proximity are not yet included.
The Coulomb expansion of the cluster sets in slowly due to the large inertia of the ions. Even after the laser pulse is switched off ($t\gtrsim2\tau$), the cluster has expanded by less than a factor 2. Our simulation shows that the charge resulting from electrons leaving the cluster (Fig.\ \ref{fig:3}(b)) is concentrated on the surface of the cluster, the ions in the inside of the cluster being well shielded by the quasi-free electrons, in agreement with the MPIC simulation \cite{brabec}. After the first ionization burst the electronic plasma frequency is given by $\omega_p^2 = N_{test} (t_1)/(\alpha R(t_1)^3) = \rho(t_1)4\pi/3\simeq0.016\,\mathrm{a.u.}\approx5\omega^2$. As electron-impact ionization produces more quasi-free electrons, $\omega_p$ grows rapidly before diminishing again as the cluster expansion sets in (Fig.\ \ref{fig:3}(c)). During the evolution, the effective field inside the cluster consisting of both the laser and the polarization field is approximately given by
\begin{eqnarray}
\label{eq:8}
\boldsymbol{F}_{eff}(t)&\approx&Re\Bigg\{\int_{\omega-\Delta\omega}^{\omega+\Delta\omega} \boldsymbol{\tilde{F}}_L(\omega')\\ 
& & \times \left(1-\frac{\omega_p^2}{\omega_p^2-\omega'^2-i\omega'\gamma}\right)e^{i\omega't}
\mathrm{d}\omega'\Bigg\}, \nonumber
\end{eqnarray}
where $\Delta \omega$ is the Fourier width caused by the temporal profile of the pulse (Eq.\ \ref{eq:2}) and $\gamma$ stands for the damping due to scattering events. $F_{eff}$ is significantly reduced compared to the bare laser field (Fig.\ \ref{fig:3}(d)) due to the combined effect of collective electron motion and electron-impact ionization. A significant resonant enhancement \cite{ditmire,saalmann} is absent. Ref. \cite{brabec} suggested laser dephasing heating (LDH), in which the phase shift between dipole moment and laser field caused by the macroscopic electric field inside the cluster, would lead to a net electron energy absorption $\int \boldsymbol{J}\boldsymbol{F}_{eff} dt$, where $\boldsymbol{J}$ stands for the electron current. However, evaluating this integral using the simulated current and field shows that heating by LDH is not sufficient in the present case. 
\begin{figure}[tb]
\includegraphics[width=8cm]{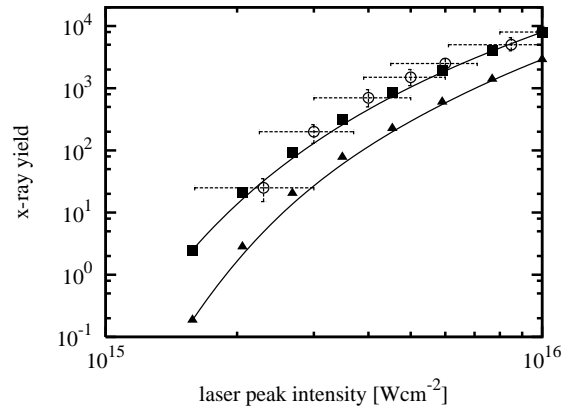}
\caption{\label{fig:4} Absolute x-ray yield as a function of the laser peak intensity. Experimental results \cite{paristhese} ($\bigcirc$), simulation results neglecting elastic electron-ion scattering ($\blacktriangle$) and including elastic electron-ion scattering ($\blacksquare$). (Lines to guide the eye).}
\end{figure}

The efficiency of heating by elastic electron-ion scattering is directly reflected in the simulated absolute x-ray yields (Fig.\ \ref{fig:4}). The latter are determined by the number of K-shell vacancies created, corrected for the mean fluorescence yield $\eta$ taken to be $\eta \approx 0.12$ \cite{bhalla} for argon with partially filled L-shell but empty M-shell. It should be emphasized that the simulation contains no freely adjustable parameter. To compare the simulation results to the experiments, an ensemble average over the spatial intensity profile of the laser beam, which is Gaussian to a good degree of approximation \cite{paristhese}, is performed. To quantify the significance of the Fermi-shuttle acceleration by repeated elastic backscattering, we performed an otherwise identical simulation with elastic electron-ion scattering switched off. In this case, for $I\gtrsim 1.5\times 10^{15}\,\mathrm{Wcm^{-2}}$ a small fraction of quasi-free electrons gains sufficient energy to produce K-shell vacancies. Their mean kinetic energy can be estimated from the potential energy of the charged-up cluster with charge $Q(t)$ (i.e.\ the monopole term of the mean effective field). However, including elastic electron-ion scattering drastically increases the x-ray yield by a factor 3 to 6. We then find surprisingly close agreement with the experimental results (Fig.\ \ref{fig:4}). 

In summary, we have analyzed the heating of the quasi-free electrons in large rare-gas clusters ($N \sim 10^5$ atoms) at moderate laser intensities ($I=10^{15}-10^{16}\mathrm{Wcm^{-2}}$). We have identified a novel, highly efficient electron heating mechanism operative at short times within a few optical cycles in terms of elastic large-angle scattering resembling the Fermi shuttle. Other processes such as heating in a plasma-resonant field are found to be less effective. In particular, the polarization of the cluster leads to a reduction rather than an enhancement of the effective field. While the surprisingly good quantitative agreement on an absolute scale with experimental data may be, in part, fortuitous, the importance of this route to fast electron acceleration appears unambiguously established. 
\begin{acknowledgments}
Work is supported by FWF SFB-16 (Austria).
\end{acknowledgments}

\end{document}